\documentclass[journal]{IEEEtran}
\usepackage{amsmath,amsfonts}
\usepackage{algorithmic}
\usepackage{array}
\usepackage[caption=false,font=normalsize,labelfont=sf,textfont=sf]{subfig}
\usepackage{textcomp}
\usepackage{stfloats}
\usepackage{url}
\usepackage{verbatim}
\usepackage{graphicx}
\usepackage{mathrsfs}
\usepackage{balance}
\usepackage[noadjust]{cite}
\usepackage{capekCommands} 
\usepackage{booktabs}

\usepackage{xcolor}
\usepackage{tikz}

\definecolor{PairedF}{RGB}{0,0,0}

\begin{document}
\title{Towards Manufacturing-Friendly Shapes in Discrete Topology Optimization}
\author{Vojtech Neuman, Miloslav Capek, \IEEEmembership{Senior Member, IEEE}, Lukas Jelinek, and Jan Spacil
\thanks{Manuscript received \today; revised \today. The Czech Science Foundation supported this work under project~\mbox{No.~21-19025M} and the Czech Technical University in Prague under project~\mbox{SGS22/162/OHK3/3T/13}.}
\thanks{V. Neuman, M. Capek, L. Jelinek and J. Spacil are with the Czech Technical University in Prague, Prague, Czech Republic (e-mails: \{vojtech.neuman; miloslav.capek; lukas.jelinek; jan.spacil\}@fel.cvut.cz).}
\thanks{Color versions of one or more of the figures in this paper are available online at http://ieeexplore.ieee.org}
}

\maketitle

\begin{abstract}
This paper addresses shape irregularity issues in discrete topology optimization algorithms where the design is created through the automated distribution of material within the design region. Graph theory is employed to derive appropriate regularity measures for any discrete optimization algorithm. Shape regularity is quantified by a set of scalar parameters ready to evaluate design choices in the form of Pareto-frontiers. Developed parameters address information regarding material usage, problematic distribution, and features that complicate manufacturing. The theory is verified by several examples demonstrating the treatment of isolated islands of materials, point connections between material segments, or material homogeneity.
\end{abstract}

\begin{IEEEkeywords}
Antennas, electromagnetic theory, inverse design, numerical methods, optimization methods, topology optimization, shape regularity.
\end{IEEEkeywords}

\section{Introduction}
\label{sec:Introduction}
\IEEEPARstart{T}{opology} optimization is a powerful and versatile tool for automated antenna design. The basic principle relies on the algorithm-driven distribution of material in a given design region~\cite{book:BendsoeTopoOpt,book:RahmatSamiiElectromagOptGenAlg}. This fundamentally different paradigm contrasts with \TR{sizing optimization~\cite{art:Levy1976ComAidDesign, Christensen2008IntroductionStructuralOptimization}}, where the antenna \TR{geometry}~\cite {book:BalanisAntTheory} is described by a set of parameters, which are then subject to an arbitrary optimization algorithm~\cite{book:MartinsEngDesOpt} to achieve the desired performance. Nevertheless, the final design must always obey a set of mechanical rules~\cite{book:LevyStructEngMicrowaveAntennas} established by a selected fabrication method~\cite{book:GrooverFundModernManufacturing}. Considering topology optimization, manufacturability represents an important underlying issue that must be managed with exceptional care. In the following, we distinguish between continuous~\cite{book:BendsoeTopoOpt} and discrete topology optimization~\cite{book:RahmatSamiiElectromagOptGenAlg}.

Continuous topology optimization~\cite{book:BendsoeTopoOpt} originates in structural analysis~\cite{book:HibbelerStructAnalysis} and has been successfully adapted for antenna design~\cite{art:Kiziltas2003TopoDesOptDielSub, art:Erentok2011TopoOptSubWave, art:Hassan2014TopoOptMetalAnt, art:Liu2016MoMTopoOptMethod, art:Tucek2023DensityBasedTopoOpt} and photonic structures~\cite{art:Christiansen2021InverseDesignPhotonics, art:Christiansen2021CompactMATLABCode, art:Schubert2022InverseDesPhotoDevices, art:Camacho2021SingleInversePhotonicStructure, art:Jenkins2023GeneralPurposeAlgorithm}.
The adjoint sensitivity formulation~\cite{book:BendsoeTopoOpt} is accompanied by filtering schemes~\cite{art:Sigmund1997OnDesignCompliant, art:Bourdin2001FiltersTopoOpt} to solve checkerboard patterns~\cite{art:Diaz1995CheckerboardPattern}. Over the years, various methods have been introduced for treating finite manufacturing tolerances~\cite{art:Sigmund2009ManufacturingTopoOpt, art:Schevenels2011RobustTopoOptSpatialErrors}, occurrences of hinges~\cite{art:Fu2014OptimizationApproachBlackAndWhite, art:Seltmann2023TopoOptCompMechanismHingeFree}, minimum feature sizes~\cite{art:Schubert2022InverseDesPhotoDevices}, problems with enclosed voids~\cite{art:Wang2020NumPerfomPoissonEnclosedVoids, art:Donoso2022NewAppSpecGraphTheory}, \TR{or general tools independent of optimization method for post-processing~\cite{art:Jenkins2023GeneralPurposeAlgorithm}}. The resulting topologies are near-binary structures that are thresholded to produce the final binary designs. However, this last step may lead to an undesired change in performance~\cite{art:Tucek2023DensityBasedTopoOpt}.

Discrete topology optimization works directly with the binary state of material distribution and utilizes nature-inspired algorithms for optimization~\cite{art:Johnson1999GenAlgMoM, art:RahmatSamii2012NatureInspOptimization}, with the newest works implementing machine learning methods~\cite{art:Jacobs2021AccModelingConvNeuralNet, art:Wang2024MachineLearningAssistPixPatch, art:Peng2025AIAssisAntOpt, inp:Bosak2025QFactorEvalAccDeepNN}. Various applications to antenna problems have been shown in several works~\cite{art:Kerkhoff2007DesignBandNotchedPlanarMonopole, art:Ethier2014AntennaShapeSynthWithoutFeedpoint, art:Yang2016SystShapeOptCharModes, art:Cismasu2014AntBandwithOpt, art:Li2022FiltAntennaCharModes}. Compared with continuous formulations, the resulting designs are significantly irregular and may obstruct manufacturing. A frequent issue arises with point connections~\cite{art:Thiel2015PointContacts}, which may be interpreted differently depending on the simulation method employed~\cite{art:Thiel2015PointContacts}. The filtering approach cannot be used as it works with continuous design variables~\cite{art:Bourdin2001FiltersTopoOpt}. One possible method for implicitly removing the problem with point connections is to use special discretization elements~\cite{art:Jayasinghe2015NonuniformOverlappingMethod,art:Mair2020EvolOptAsyPixAntennas} or to introduce auxiliary logic in optimization~\cite{art:Jin2010HybridRealBinaryPSO}. However, nontrivial elements prevent the utilization of fast evaluation schemes based on exact reanalysis~\cite{Ohsaki2011, art:Capek2019ShapeSyntTopoSens, inp:Wang2022BinaryTopoModel, art:Jiang2022PixAntenanOptimization, art:Budhu2023FastAccOptMetasurface}. Regularity and manufacturability are usually solved only as a post-processing step (\eg,{} slightly shifting problematic nodes or removing insignificant elements~\cite{art:Ethier2014AntennaShapeSynthWithoutFeedpoint}). The requirement of post-processing prolongs design time and introduces potential undesired changes in performance.

\TR{This article introduces manufacturability as a standalone metric quantifiable by a set of scalar figures of merit, evaluating performance with respect to the regularity of the structure and the presence of local problematic spots, such as point connections~\cite{art:Thiel2015PointContacts}. The obtained information can be used either as a post-processing detection mechanism or directly during optimization as additional criteria, similarly to works~\cite{art:Cismasu2014AntBandwithOpt, art:Seltmann2023TopoOptCompMechanismHingeFree}. The described method can be used in any optimization paradigm where the structure is encoded as a binary vector, such as in a genetic algorithm~\cite{art:Johnson1999GenAlgMoM} or, more recently, in machine learning~\cite{art:Jacobs2021AccModelingConvNeuralNet}. As a continuation of previous authors' works~\cite{art:Capek2019ShapeSyntTopoSens, art:Capek2021MemeticSchemeTopoOptI, art:Capek2021MemeticSchemeTopoOptII, art:Kadlec2025MultiObjMemeticAlg}, the presented theory is demonstrated on the optimization of electrically small antennas by means of a memetic algorithm~\cite{art:Capek2021MemeticSchemeTopoOptI}, where the optimized metric consists of Q-factor, reflection coefficient, and, in addition, regularity parameters. Nevertheless, as mentioned above, it could be used in connection with an arbitrary discrete scheme.}

The initial idea is based on the representation of a discretized topology by graph theory. Previous works used graphs as simplified shapes for topology optimization~\cite{art:Giger2006EvolutionaryTrussTopoOpt, art:Jie2021NovelWeightGraphRepresentation}. Here, we assume that each discretization element corresponds to an individual graph element. This approach enables the introduction of powerful graph tools that can be used to deal with manufacturability problems (\eg{}, enclosed voids~\cite{art:Donoso2022NewAppSpecGraphTheory}).

The rest of this article is organized as follows. Section~\ref{sec:MotivationExample} introduces the main idea using an example featuring two antennas. Next, graph theory is introduced in Section~\ref{sec:GrapthTheoryAndMoM}. Section~\ref{sec:Metrics} shows the utilization of graph theory on the definition of regularity parameters. Section~\ref{sec:Examples} provides an additional set of examples. The article concludes in Section~\ref{sec:Conclusion}.

\section{Regularity Problem}
\label{sec:MotivationExample}
Manufacturability issues arising in discrete topology optimization are introduced on the example of an antenna designed by a memetic algorithm~\cite{art:Capek2021MemeticSchemeTopoOptI}. The algorithm proceeds by forming current paths by either adding or removing the individual basis functions, effectively introducing either a short or virtual open between two triangles. Details concerning electromagnetic modeling used throughout this article are summarized in~\cite{art:Jelinek2017OptCurrArbitShapes}.

The regularity issue is presented on a specific example concerning the design of an electrically small antenna within the given design region $\varOmega$ with electrical size $ka = 0.7$, with $k$ being a wavenumber and $a$ the radius of the sphere fully circumscribing the radiating body. The requirements maximize the fractional bandwidth with reflection coefficient $\Gamma$ on design frequency $\left\vert\Gamma\right\vert \leq 3.1\cdot10^{-2}$, equivalent to a return loss~\cite{book:PozarMicroEng} below $-30\,$dB. The design space is a rectangle of size $100\,$mm$ \times 50\,$mm, consisting of a copper sheet with conductivity $\sigma = 5.96\cdot10^{7}$\,S/m and low loss dielectric substrate Astra MT77~\cite{dat:ASTRAMT77} with dielectric constant $\varepsilon_\T{r} = 3$, dissipation factor $\tan{\delta} = 0.0017$ and a height of $1.524\,$mm. The frequency is $f_0 = 597.47\,$MHz.

\begin{figure}
\centering
\includegraphics[width=0.99\columnwidth]{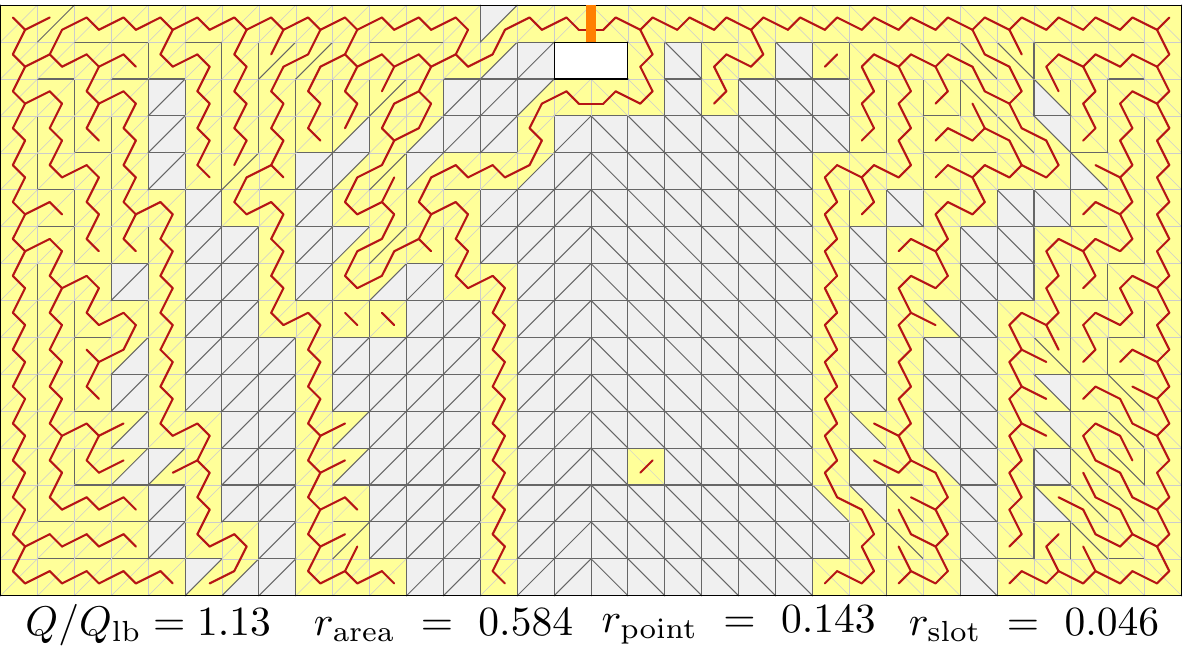}
\caption{A solution to problem~\eqref{eq:TopoOptUnrestricted} found by the algorithm from~\cite{art:Capek2021MemeticSchemeTopoOptI}. The electrical size of the design domain is~$ka=0.7$. The orange line represents the position of the discrete feeding port~\cite{gibson2021moments}. The value~$r_\T{area}$ measures the used area, and parameters~$r_\T{point}$ and~$r_\T{slot}$ quantify the presence of point connections and infinitesimal slots, respectively. The exact definition is provided in later sections.}\label{pic:TopoOptExampleUnrestricted}
\end{figure}

The multi-objective optimization problem reads
\begin{equation}
\begin{split}
\underset{\Gvec}{\T{minimize}} \quad & \left\{\frac{Q}{Q_\T{lb}},\left\vert\Gamma\right\vert\right\}\ \\
\end{split}
\label{eq:TopoOptUnrestricted}
\end{equation}
where $\Gvec$ is the optimization variable encoding the presence of the basis functions, $Q$ and $Q_\T{lb}$ represent the Q-factor and its corresponding lower bound~\cite{art:Capek2017MinAntQ, art:Gustafsson2016AntCurrOpt}. \TR{Problem~\eqref{eq:TopoOptUnrestricted} can be solved with an arbitrary solver working with binary genes~$\Gvec$. This work uses the algorithm described in~\cite{art:Capek2021MemeticSchemeTopoOptI}, with the found design shown in Fig.~\ref{pic:TopoOptExampleUnrestricted}.} The normalized Q-factor is $Q/Q_\T{lb} = 1.13$.

\begin{figure}
\centering
\includegraphics[scale=1]{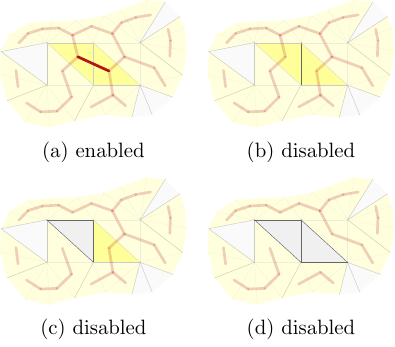}
\caption{State of the $n$-th basis function and its corresponding triangles. Red lines represent enabled basis functions, yellow triangles are conductive material, and gray triangles are void. (a) Basis function and both triangles are enabled. (b) The basis function is disabled, and both triangles are enabled. (c) The basis function is disabled, one triangle is enabled, and the second is disabled. (d) The basis function and both triangles are disabled. The current can flow between both triangles only in case (a).}\label{pic:ElementBF}
\end{figure}

\begin{figure}
\centering
\includegraphics[scale=1]{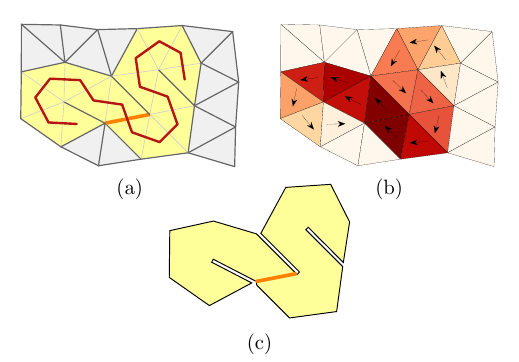}
\caption{Topology formed by enabling or disabling the current flow between two adjacent triangles. Each red segment represents two connected triangles. The orange line represents a discrete feeding port. (a) A given topology with material-filled triangles in yellow. (b) A current density resulting from the connection of the delta gap in the middle of the current path. (c) Topology with physical slots approximating infinitesimal slots in panel (a). }\label{pic:SlotExample}
\end{figure}

Figure~\ref{pic:TopoOptExampleUnrestricted} represents the copper layer within design region $\varOmega$ discretized into a set of triangles interconnected by means of \ac{RWG} basis functions~\cite{art:Rao1982EleScattSurf}. Figure~\ref{pic:ElementBF} illustrates the principle of current path generation on the example of the $n$-th basis function. Figure~\ref{pic:ElementBF}(a) shows the $n$-th basis function as enabled and, therefore, a current can flow between the two highlighted triangles. In the other panels, the basis function is disabled, and the current cannot flow between the two triangles even though there are still triangles present in cases (b) and (c). Dielectric support of the rectangular design domain is not subject to optimization.
Figure~\ref{pic:SlotExample} shows the effect of vector $\Gvec$ on the current flow excited by a discrete feeding port. It also illustrates how a situation from Fig.~\ref{pic:ElementBF}(b) translates to the final design. Two enabled triangles with their corresponding basis function removed represent a situation where the structure contains an infinitesimal slot~\cite{art:Capek2019ShapeSyntTopoSens}. For the sake of manufacturing, such slots are made with a finite width~\cite{art:Capek2019ShapeSyntTopoSens}.

A visual inspection of Fig.~\ref{pic:TopoOptExampleUnrestricted} reveals that the optimized design suffers from several deficiencies, of which the most salient ones are summarized in Fig.~\ref{pic:TopoOptManufactureProblems} and listed below:
\begin{figure}
\centering
\includegraphics[width=0.4\textwidth]{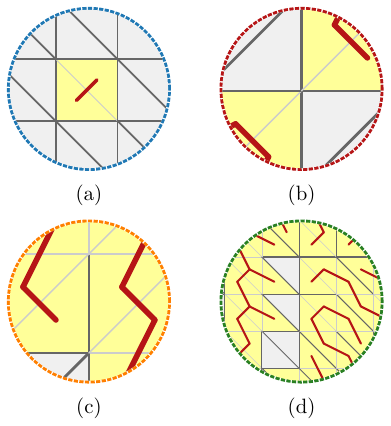}
\caption{A sketch of the most salient issues related to discrete topology optimization. The depicted features are taken from Fig.~\ref{pic:TopoOptExampleUnrestricted}. (a) Isolated metal island, (b) point connection, (c) infinitesimal slots, (d) rapidly changing design variable.}
\label{pic:TopoOptManufactureProblems}
\end{figure}
\begin{enumerate}
\item The first issue concerns the isolated metal islands, shown in Fig.~\ref{pic:TopoOptManufactureProblems}(a), which are elements that do not usually affect the performance of the resulting structure considerably and can be omitted. Isolated islands are not a manufacturing concern when the antenna is etched on a board. However, they can pose an issue when the considered design is intended to be self-supporting, such as in the case of a dielectric lens~\cite{art:Christiansen2021CompactMATLABCode, art:Christiansen2021InverseDesignPhotonics}.
\item Another issue is the presence of point connections~\cite{art:Jayasinghe2012SimpleDesignMultiBandPatch, art:Jayasinghe2015NonuniformOverlappingMethod, art:Mair2020EvolOptAsyPixAntennas, art:Thiel2015PointContacts}, where one point shared by several triangles does not physically connect those triangles, see Fig.~\ref{pic:TopoOptManufactureProblems}(b). This introduces ambiguity in the simulation as different methods treat point connections in different fashions~\cite{art:Thiel2015PointContacts}, and uncertainty whether the point in manufactured design would be connected or disconnected.
\item The problem depicted in Fig.~\ref{pic:TopoOptManufactureProblems}(c) has already been mentioned above. This is the usual drawback of fast update methods~\cite{art:Capek2019ShapeSyntTopoSens, inp:Wang2022BinaryTopoModel} based on the Sherman-Morrison-Woodbury formula~\cite{book:GolubMatrixComp}. In the context of this work, this issue is referred to as an infinitesimal slot and shares similar manufacturing problems to point connections. Its occurrence has to be replaced with a physical slot~\cite{art:Capek2019ShapeSyntTopoSens}, see Fig.~\ref{pic:SlotExample}, of finite width.
\item Lastly, Fig.~\ref{pic:TopoOptManufactureProblems}(d) illustrates the problem of frequent changes of the optimization variable causing irregularities in the design~\cite{art:Cismasu2014AntBandwithOpt, art:Capek2019ShapeSyntTopoSens}. In continuous topology optimization, this issue is called the checkerboard pattern~\cite{art:Diaz1995CheckerboardPattern}. Frequent material changes demand manufacturing precision and sophisticated computer modeling.
\end{enumerate}
Handling the listed drawbacks can worsen performance and be time-consuming when done manually. Manufacturing problems resulting from discrete topology optimization are influenced by several factors, such as the manufacturing method, whether it be subtractive or additive, and electrical size, which directly affects minimum feature sizes~\cite{art:Jenkins2023GeneralPurposeAlgorithm}. The finite precision of tools, which causes random errors~\cite{art:Schevenels2011RobustTopoOptSpatialErrors}, necessitates robust and unambiguous designs.

This work addresses the aforementioned issues by incorporating an additional metric $R$ considering the regularity onto the objective function~\eqref{eq:TopoOptUnrestricted}. The modified optimization problem reads
\begin{equation}
\begin{split}
\underset{\Gvec}{\T{minimize}} \quad & \left\{\frac{Q}{Q_\T{lb}},\left\vert\Gamma\right\vert,R\right\}
\end{split}
\label{eq:TopoOptRestricted}
\end{equation}
where the regularity metric $R$ is defined as
\begin{equation}
R = w_1r_\T{area} + w_2r_\T{point} + w_3r_\T{slot},
\end{equation}
with $w_1$, $w_2$ and $w_3$ being weights balancing the individual goals which are $r_\T{area}$ representing a normalized metallized area, $r_\T{point}$ evaluating the number of elements sharing one point and $r_\T{slot}$ providing information on the presence of infinitesimal slots. The exact formulas for the introduced regularity parameters are defined later in Section~\ref{sec:Metrics}.

\begin{figure}
\centering
\includegraphics[width=0.99\columnwidth]{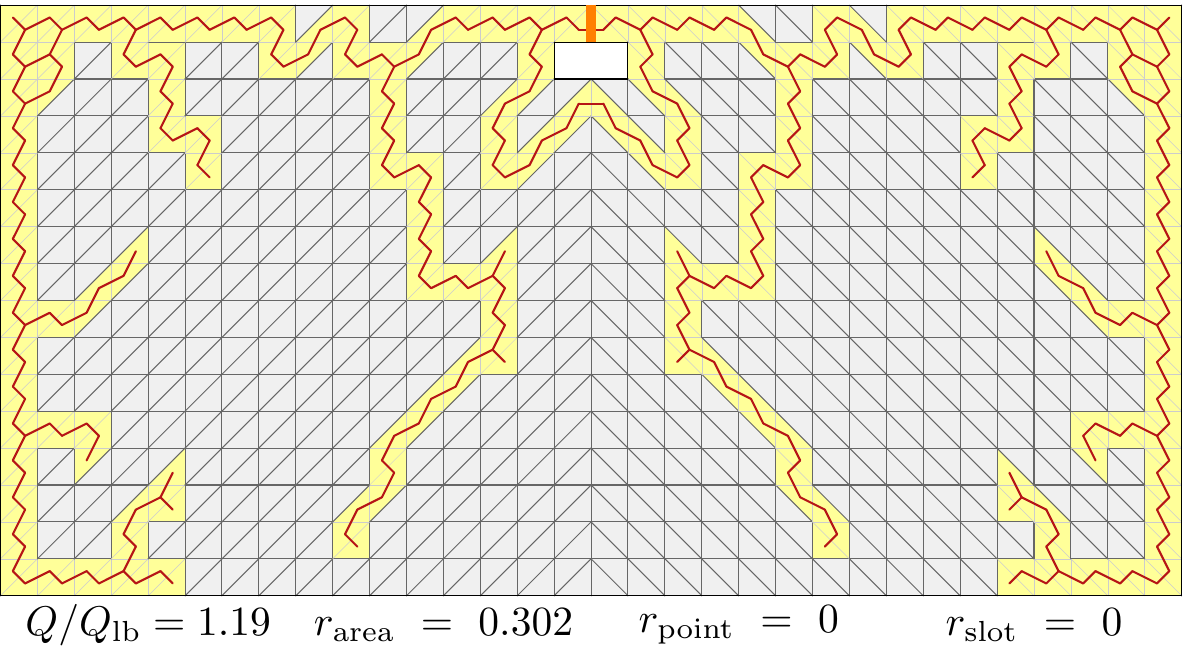}
\caption{A solution to problem~\eqref{eq:TopoOptRestricted} found by the algorithm from~\cite{art:Capek2021MemeticSchemeTopoOptI}. The electrical size of the design domain is~$ka=0.7$. The orange line represents the position of the discrete feeding port. The value~$r_\T{point}$ quantifies the presence of point connections.}\label{pic:TopoOptExampleRestricted}
\end{figure}

The design obtained from the optimization problem~\eqref{eq:TopoOptRestricted} is shown in Fig.~\ref{pic:TopoOptExampleRestricted} with the corresponding values of physical metrics and regularity parameters. The return loss is again below $-30\,$dB. The normalized Q-factor is slightly higher than in the case of Fig.~\ref{pic:TopoOptExampleUnrestricted}. However, the structure itself is regular and easily manufacturable without further post-processing or compromises. The mirror symmetry in the optimization variable $\Gvec$ was employed to further increase the overall regularity of the result~\cite{9769019}.

\begin{figure}
\centering
\includegraphics[width=0.99\columnwidth]{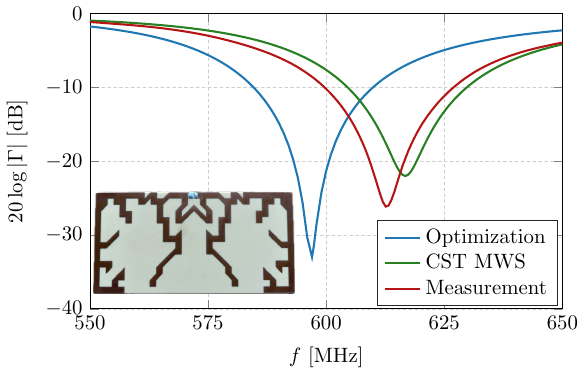}
\caption{Comparison between initial design (blue) and measured prototype (red). The green curve represents the auxiliary simulation from CST Microwave Studio, including a simplified model of the realized feeder. The inset shows the manufactured antenna.}\label{pic:TopoOptPrototypeComparison}
\end{figure}

The proposed design was manufactured with printed circuit board technology and measured, see Appendix~\ref{sec:ManuMeas} for details on the measurement configuration and feeding structure. The comparison between the simulation and measurement is shown in Fig.~\ref{pic:TopoOptPrototypeComparison}. The fractional bandwidth resulting from the simulation corresponds well to the measured prototype. The central frequency is slightly shifted towards higher frequencies. Inconsistency is attributed to the finite precision of numerical modeling, together with the limited ability of the realization of a discrete port by realistic feeding. This was verified by a secondary simulation conducted in CST Microwave Studio~\cite{web:CST} with two waveguide ports, see Appendix~\ref{sec:ManuMeas}.

The fractional bandwidth (FBW) evaluated at $-10\,$dB level is compared in Tab.~\ref{tab:FBW}.
\begin{table}[ht]
\centering
\renewcommand{\arraystretch}{1.5} 
\caption{Fractional bandwidth comparison.}\label{tab:FBW}
\begin{tabular}{lccc}
\toprule
& Optimization & CST MWS & Measurement \\
FBW  & $4.2\,$\% & $4.1\,$\% & $4.4\,$\% \\
\bottomrule
\end{tabular}
\end{table}

The rest of this article aims to develop the theory for regularity parameters and to show how to deal with multiple regularity parameters during the design phase.

\section{Topology Representation by Graph Theory}
\label{sec:GrapthTheoryAndMoM}
This section introduces the fundamental theoretical background through an example of the design region discretized into polygons. 

\begin{figure}
\centering
\includegraphics[scale=1]{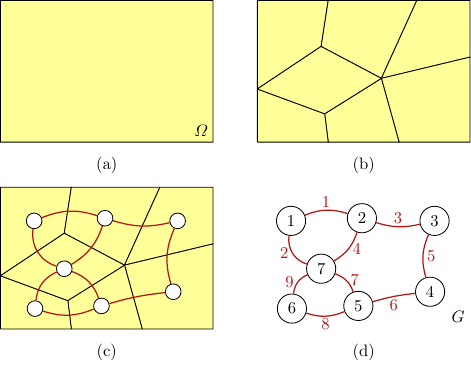}
\caption{Graph representation of the optimization region. (a) Arbitrary region~$\varOmega$ to be discretized, (b) discretization of~$\varOmega$ into a set of arbitrary elements, (c) highlighted connections between important sets, (d) graph representation of~$\varOmega$.}\label{pic:Mesh2Graph}
\end{figure}

The discretization elements constitute a set of vertices~$\mathscr{V}$. A polygon edge shared by two adjacent polygons~$v_n, v_m\in \mathscr{V}$ forms graph edge~$e_j \in \mathscr{E}$, and both sets form graph~$G = \left(\mathscr{V}, \mathscr{E}\right)$. Figure~\ref{pic:Mesh2Graph} depicts the situation for a rectangle discretized into arbitrary polygons and visualizes the transition process from the original continuous object to graph~$G$.

\begin{figure}
\centering
\includegraphics[]{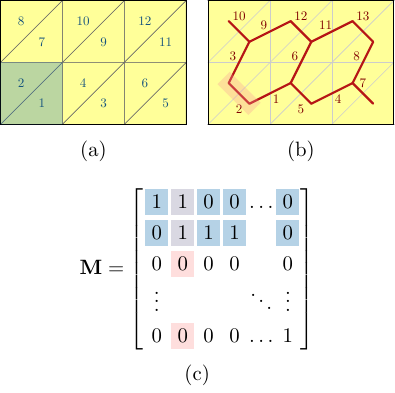}
\caption{Visualization of graph vertices $\mathscr{V}$ (triangles) and graph edges $\mathscr{E}$ (triangles edges). (c) The red and blue colors highlight the relation between both sets in the incidence matrix~$\Mmat$.}\label{pic:GraphMatrices}
\end{figure}

\begin{figure}
\centering
\includegraphics[scale=1]{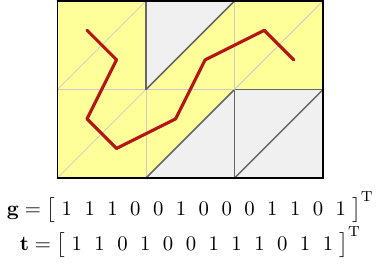}
\caption{Example of triangle genes $\Gvec$ and $\Tvec$ and a structure corresponding to it. Enumeration of individual basis functions and triangles is the same as in Fig.~\ref{pic:GraphMatrices}.}\label{pic:GeneExample}
\end{figure}

Material manipulation in discrete topology optimization directly translates into enabling or disabling graph vertices and their corresponding edges. Encoding the presence of discretization elements in binary values defines a logical vector
\begin{equation}
t_m = 
  \begin{cases}
    \displaystyle\, 1 & \text{if the $m$-th vertex is present}, \\
    \displaystyle\, 0 & \text{otherwise}.
  \end{cases}
\end{equation}
Similarly, the presence of graph edges between vertices establishes a logical vector
\begin{equation}
g_n = 
  \begin{cases}
    \displaystyle\, 1 & \text{if the $n$-th edge is present}, \\
    \displaystyle\, 0 & \text{otherwise}.
  \end{cases}
\end{equation}
Vectors $\Tvec$ and $\Gvec$ are not independent. The presence of vertices and edges is related to the incidence matrix $\Mmat$, see Appendix~\ref{sec:GraphMatrices}, and the relation is defined as
\begin{equation}
\Tvec = \B{\Mmat\Gvec}, \label{eq:Edge2Vert}
\end{equation}
with the element-wise Boolean rounding operator
\begin{equation}
\B{x} = 
  \begin{cases}
    \displaystyle\, 0 & x = 0, \\
    \displaystyle\, 1 & \text{otherwise}.
  \end{cases}\label{eq:BooleanRounding}
\end{equation}
which maintains all values of both vectors binary\footnote{Since matrix $\Mmat$ is rank-deficient, several vectors~$\Gvec$ can produce the same vector~$\Tvec$.}. To further illustrate the meaning of the incidence matrix, Fig.~\ref{pic:GraphMatrices} shows its application to triangles that constitute a set of vertices and their shared edges, providing a set of graph edges.
Vectors $\Gvec$ and $\Tvec$ define the current shape, as shown in Fig.~\ref{pic:GeneExample}, and can be used to extract information about the topology from the graph matrices.

\section{Regularity Parameters}
\label{sec:Metrics}
The problems with manufacturability described in Section~\ref{sec:MotivationExample}  are quantified by a set of so-called regularity parameters~$r$.

The first shape parameter refers to the isolated islands and is quantified by relative area~$r_\T{area}$. Adding this parameter to the goal function forces the optimizer to omit parts that do not significantly contribute to the optimized metric.

The treatment of slots and point connections is based on identifying them in the topology and minimizing their number. This is quantified by the parameters $r_\T{slot}$, respectively $r_\T{point}$, which utilize Boolean operations for detecting these defects and acquiring their count.

Lastly, the frequent changes in design variable are addressed by the homogeneity parameter $r_\T{hom}$. Homogeneity is evaluated by checking the neighborhood of each element and calculating the average state of its adjacent optimization variables.

All four parameters~$r_\T{area}$, $r_\T{slot}$, $r_\T{point}$, and $r_\T{hom}$ are normalized to a span interval $\left[0,1\right]$ and increase when the phenomenon they are describing increases. For example, a high value of $r_\T{area}$ represents a structure primarily filled with material.

The formal definitions of the aforementioned parameters are detailed in the following subsections. In the rest of the article, the discretization elements are triangles, and \ac{RWG} basis functions serve as optimization variables. However, the theory can be applied to an arbitrary scheme where design optimization can be represented as a graph.

\subsection{Conducting Area}
The relative area is defined as
\begin{equation}
r_\T{area}(\Gvec) = \frac{\M{a}^\trans \B{\Mmat\Gvec}}{\M{a}^\trans\B{\Mmat\Gvec_0}} = \frac{\M{a}^\trans\Tvec}{\left\Vert \M{a}\right\Vert_1},
\label{eq:RelativeAreaTriaT}
\end{equation}
where $\M{a} \in \mathbb{R}^{T\times 1}$ is a vector containing the surface area of each discretization element (triangle), $\Gvec_0$ is the vector of the design region fully spanned by material, and the denominator is the total area of the optimization region. A similar idea has already been adopted in~\cite{art:Cismasu2014AntBandwithOpt} for a regular grid.

\subsection{Infinitesimal Slots}
\label{sec:SlotMetric}
The infinitesimal slot arises when two adjacent triangles are enabled while the basis function connecting them is disabled. This leads to a definition of the slot parameter
\begin{equation}
r_\T{slot}\left(\Gvec\right) = \frac{\left\Vert \Gvec \oplus  \widetilde{\Gvec}\right\Vert_1}{B -\left\lfloor\dfrac{T}{2}\right\rfloor},\label{eq:rSlot}
\end{equation}
where $\oplus$ represents the XOR operation.

Relation~\eqref{eq:rSlot} utilizes the properties of graph matrices by calculating the modified gene
\begin{equation}
\widetilde{\Gvec} = \neg\B{\Mmat^\trans\B{\Mmat\Gvec}-2\Gvec_0}, \label{eq:BasisVectorWithSlots}
\end{equation}
where all infinitesimal slots are removed. The expression proceeds by first converting vector $\Gvec$ to vector $\Tvec$ by means of relation~\eqref{eq:Edge2Vert}, resulting in active triangles. Multiplying the triangle gene with the transposed matrix $\Mmat^\trans$ returns a vector containing information on how many active triangles are adjacent to a given basis function. As enabled basis functions must always consist of two active triangles, subtracting the expression $2\Gvec_0$ and subsequently applying the Boolean rounding operator returns the logically inverted gene with slots filled. The final application of Boolean negation $\neg$ inverts all binary values to the correct level. Comparing the original and reconstructed vectors leads to the conclusion that
\begin{equation}
\text{number of slots} = \left\Vert \Gvec \oplus \widetilde{\Gvec}\right\Vert_1.\label{eq:NoSlot}
\end{equation}
The vector inside the norm is a gene that contains only enabled basis functions located exactly at the positions of the slots.

Normalization in~\eqref{eq:rSlot} is provided by the maximum number of slots in a given topology. This can be done with graph matching~\cite{book:BollobasModernGraphTheory}, see Appendix~\ref{sec:GraphMatching}. Perfect matching~\cite{book:BollobasModernGraphTheory} has a cardinality (\ie{}, a number of active edges)
\begin{equation}
\left\vert \mathscr{M}\right\vert = \frac{T}{2}.
\end{equation}
Using this knowledge, the upper bound on the number of slots can be estimated as
\begin{equation}
\max_\Gvec{\left\Vert \Gvec \oplus  \widetilde{\Gvec}\right\Vert_1} \approx B - \left\lfloor\frac{T}{2}\right\rfloor,
\end{equation}
where the floor operator solves the nonexistence of a perfect matching for an odd number of triangles. Putting both parts together constitutes the slot metric~\eqref{eq:rSlot}.

\subsection{Point Connections}
A point connection exists when two triangles share only one common point~\cite{art:Thiel2015PointContacts}, and \ac{MoM}, based on \ac{RWG} basis functions~\cite{art:Rao1982EleScattSurf}, does not consider these two triangles to be conductively connected. The occurrence of point connections is quantified by 
\begin{equation}
r_\T{point}\left(\Gvec\right) = \frac{\left\Vert\OP{H}\left\{\M{p} - 2\M{p}_0\right\}\right\Vert_1 }{N - \left\Vert\OP{H}\left\{-\M{p}\right\}\right\Vert_1}.\label{eq:rPoint}
\end{equation}
Evaluation of~\eqref{eq:rPoint} requires vector~$\M{p}$ defined as

\begin{equation}
\M{p} = \Mmat_\T{nt}\Tvec - \Mmat_\T{nb}\widetilde{\Gvec},
\end{equation}

where $\widetilde{\Gvec}$ is a gene defined by relation~\eqref{eq:BasisVectorWithSlots}, and where $\Mmat_\T{nt}$ and $\Mmat_\T{nb}$ are additional incidence matrices defined in Appendix~\ref{sec:GraphMatrices}, which relate the triangles and basis functions to their incident nodes. Notice that vector~$\widetilde{\Gvec}$ can be reused when both~\eqref{eq:rSlot} and~\eqref{eq:rPoint} are evaluated.

\begin{figure}
\centering
\includegraphics[scale=1]{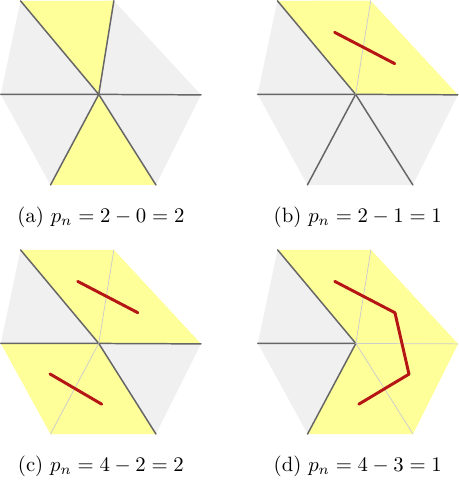}
\caption{Evaluation of the~$n$-th component of point connection vector~$\M{p}$ in different situations. The investigated node lies in the center. Number $p_n$ is equal to the number of red basis functions subtracted from the yellow triangles.}\label{pic:MetricPointTR}
\end{figure}

The meaning of vector~$\M{p}$ is shown in Fig.~\ref{pic:MetricPointTR} for the $n$-th point. The number $p_n$ is evaluated by counting the enabled incident triangles and subtracting the enabled incident basis functions. It is always greater than or equal to zero. Cases (a) and (c) are undesired, and the corresponding value~$p_n$ is greater than one.
Using this knowledge, it is possible to calculate the number of problematic points (numerator of~\eqref{eq:rPoint})
\begin{equation}
\text{number of problematic nodes} = \left\Vert\OP{H}\left\{\M{p} - 2\M{p}_0\right\}\right\Vert_1,
\end{equation}
where $\M{p}_0$ is a vector full of ones ($p_{n} = 1,\forall n$) and $\OP{H}\left\{-\right\}$ is an element-wise Heaviside operator
\begin{equation}
\OP{H}\left\{x\right\} =
\begin{cases}
    \displaystyle 1 & x \geq 0, \\
    \displaystyle 0 & \text{otherwise}.
\end{cases}
\end{equation}

The point connection metric could be normalized to the total number of nodes $N$, however, it is convenient to subtract all nodes for which $p_n = 0$. Such nodes are fully encircled by metal or a vacuum. The number of such nodes is equal to~$\left\Vert\OP{H}\left\{-\M{p}\right\}\right\Vert_1$. This reduced number is used as the denominator in~\eqref{eq:rPoint}.

\subsection{Metal Distribution}
\label{sec:MetalDistrTR}
\begin{figure}
\centering
\includegraphics[scale=1]{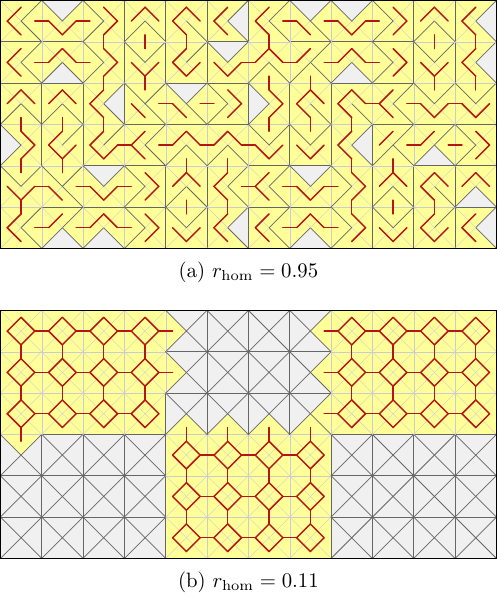}
\caption{Comparison of the homogeneity parameter~\eqref{eq:RegMetric} for two designs.}\label{pic:RegularityExample}
\end{figure}

The homogeneity of the design is quantified using
\begin{equation}
r_\T{hom}\left(\Gvec\right) = \frac{B - \left\Vert2\Hmat\Gvec- \Gvec_0\right\Vert_1}{B - \sum\limits_{n}\min\limits_{\Gvec}\left\vert2\Hmat_n\Gvec - 1\right\vert},
\label{eq:RegMetric}
\end{equation}
where $\Hmat$ is the homogeneity matrix defined in Appendix~\ref{sec:GraphMatrices} and $\Gvec_0$ is a basis function vector corresponding to a fully-filled design region ($g_n = 1,\forall n$). Figure~\ref{pic:RegularityExample} shows the value of~\eqref{eq:RegMetric} for two different topologies. It is observed that rapid changes in design variables lead to high values of ~$r_\T{hom}$.

\begin{figure}
\centering
\includegraphics[scale=1]{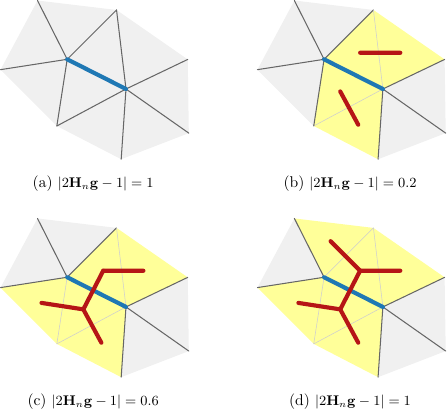}
\caption{Demonstration of homogeneity $\left\vert 2\Hmat_n\Gvec-1\right\vert$ applied to the $n$-th basis function, highlighted by the blue edge crossing it. As seen above: (a) $n$-th basis function disabled as are all neighboring basis functions -- shares the highest homogeneity with (d); (b) $n$-th basis function disabled but two neighboring basis functions enabled -- poor homogeneity; (c) $n$-th basis function enabled as are three of the four neighboring basis functions -- acceptable homogeneity; (d) all basis functions locally enabled -- shares the highest homogeneity with (a).}\label{pic:MetricRegularity}
\end{figure}

A key component of  parameter~\eqref{eq:RegMetric} is the investigation of the neighborhood of each element, which is provided by the value
\begin{equation}
\left\Vert2\Hmat\Gvec - \Gvec_0\right\Vert_1 = \sum_{n}\left\vert2\M{H}_n\Gvec - 1\right\vert, \label{eq:rRegExpression}
\end{equation}
where $\Hmat_n$ is the $n$-th row of the homogeneity matrix. The effect of relation~\eqref{eq:rRegExpression} is depicted for an $n$-th summation member in Fig.~\ref{pic:MetricRegularity}, showing several different material distributions in elements neighboring the $n$-th basis function. Figure~\ref{pic:MetricRegularity}(a) and Fig.~\ref{pic:MetricRegularity}(d) show the monotonous distribution, and the value of expression~\eqref{eq:rRegExpression} is maximal. In contrast, Fig.~\ref{pic:MetricRegularity}(b) and Fig.~\ref{pic:MetricRegularity}(c) show situations where the neighborhood of the $n$-th element is not homogeneous.

The normalization of the homogeneity parameter is obtained using
\begin{equation}
\min_{\Gvec}\left\Vert2\Hmat\Gvec - \Gvec_0\right\Vert_1 \leq \sum_{n}\min_{\Gvec}\left\vert2\Hmat_n\Gvec - 1\right\vert, \label{eq:rRegNormalization}
\end{equation}
where the right side is significantly easier to evaluate.

\section{Properties and Applications}
\label{sec:Examples}
The series of examples shows how the introduced regularity parameters can be utilized to generate manufacturing-friendly optimal shapes. The methodology for multi-objective optimization introduced in work~\cite{art:Kadlec2025MultiObjMemeticAlg} is adopted for dealing with multiple physical metrics and regularity parameters.

\subsection{Topology Evaluation by Regularity Parameters}
\begin{figure}
\centering
\includegraphics[width=0.48\textwidth]{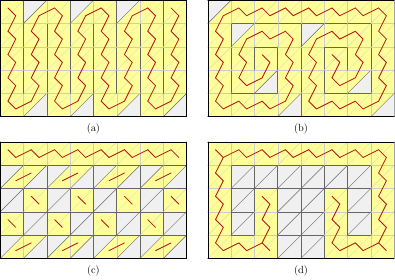}
\caption{Various topologies highlighting manufacturing defects. The values of the regularity parameters are shown in Tab.~\ref{tab:RegParam}.}\label{pic:RegParamExample}
\end{figure}

Four simple topologies shown in Fig.~\ref{pic:RegParamExample} are considered for comparison by the regularity parameters. Figure~\ref{pic:RegParamExample}(a) represents a tightly-spaced meandered antenna. Though the numerical modeling would produce an expected result, the individual arms cannot be manufactured without a post-processing step introducing the finite-width slots. A similar situation occurs for Fig.~\ref{pic:RegParamExample}(b) with a Palmier-like shape~\cite{art:Jelinek2018RadEffCostReson}. Both figures experience significant values in metric $r_\T{slot}$, see Tab.~\ref{tab:RegParam}, quantifying problematic infinitesimal slots. The checkerboard pattern~\cite{art:Diaz1995CheckerboardPattern} in Fig.~\ref{pic:RegParamExample}(c) is captured by the increased value of $r_\T{point}$. Figure~\ref{pic:RegParamExample}(d) is a shape without any of the problems mentioned above and limited only by manufacturing tolerances.
The regularity parameters for the presented topologies are listed in Tab.~\ref{tab:RegParam}.

\begin{table}[ht]
\centering
\renewcommand{\arraystretch}{1.5}
\caption{Regularity parameters of various topologies.}\label{tab:RegParam}
\begin{tabular}{ccccc}
\toprule
Design & $r_\T{area}$ & $r_\T{slot}$ & $r_\T{point}$ & $r_\T{hom}$ \\ 
\midrule
Fig.\ref{pic:RegParamExample}(a) & 0.91 & 0.33 & 0      & 0.75 \\
Fig.\ref{pic:RegParamExample}(b) & 0.91 & 0.28 & 0      & 0.72 \\
Fig.\ref{pic:RegParamExample}(c) & 0.6 & 0.06 & 0.39 & 0.49 \\
Fig.\ref{pic:RegParamExample}(d) & 0.6 & 0      & 0      & 0.43 \\ 
\bottomrule
\end{tabular}
\end{table}

\subsection{Computational Complexity}
\begin{figure}
\centering
\includegraphics[width=0.48\textwidth]{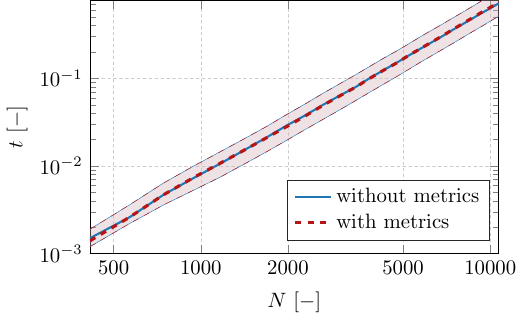}
\caption{Comparison of computational complexity of fitness functions~\eqref{eq:CompComplexity0} and~\eqref{eq:CompComplexity1} depending on growth in the number of unknowns $N$. Thick blue and red lines represent median evaluation time. The upper and lower thin curves highlight the 90th and 10th percentiles, respectively. }\label{pic:CompComplexity}
\end{figure}

Including regularity parameters in the optimization introduces additional computational demands on the fitness function evaluation. This example compares two functions, one consisting of a single physical metric
\begin{equation}
f =  \frac{Q}{Q_\T{lb}},\\
\label{eq:CompComplexity0}
\end{equation}
and a second, adding all the introduced regularity parameters
\begin{equation}
\tilde{f} = \frac{Q}{Q_\T{lb}} + r_\T{area} + r_\T{hom} + r_\T{point} + r_\T{slot}.\\
\label{eq:CompComplexity1}
\end{equation}
The evaluation times are computed for both functions for an increasing number of discretization elements. The results are shown in Fig.~\ref{pic:CompComplexity}.
Medians are compared together with the confidence intervals of the 10th and 90th percentiles. It can be seen that both distributions are almost identical.

This is the result of the effective implementation allowed by the sparsity of the graph operators. To better illustrate this property, the last sample in Fig.~\ref{pic:CompComplexity} has $N = 10710$ unknowns. Considering the homogeneity matrix $\Hmat$, the number of nonzero entries is $53190$, which is less than half 
per mil of the total number of matrix entries.

\subsection{Trade-Off Between Q-factor and Relative Area}
A well-known result of small antenna theory is the inverse cubic dependence of the Q-factor on electrical size~\cite{book:VolakisSmallAnt}
\begin{equation}
Q \propto \frac{1}{\left(ka\right)^3}. \label{eq:Q}
\end{equation}

\begin{figure}
\centering
\includegraphics[width=0.48\textwidth]{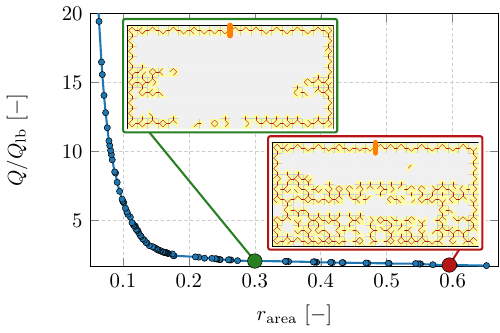}
\caption{Pareto frontier resulting from multi-objective optimization~\eqref{eq:TopoOptQnormArel}. The blue curve represents the fit between individual solutions. The red and green dots, along with their respective insets, represent selected solutions to the problem. The discrete port feeding of both designs is highlighted with the orange line in the center of the top side.}\label{pic:TopoOptParetoQnormArel}
\end{figure}

This example, similar to~\cite{art:Cismasu2014AntBandwithOpt}, deals with the trade-off between the area used for the antenna and the Q-factor
\begin{equation}
\begin{split}
\underset{\Gvec}{\T{minimize}} \quad & \left\{\frac{Q}{Q_\T{lb}},r_\T{area}\right\}.
\end{split}
\label{eq:TopoOptQnormArel}
\end{equation}
The resulting Pareto frontier is depicted in Fig.~\ref{pic:TopoOptParetoQnormArel}. Structures with normalized Q-factors lower than $Q/Q_\T{lb} = 1.5$ span more than 50\,\% of the design area\footnote{At a certain level of utilized area, the Q-factor starts to increase again.} ($r_\T{area} > 0.5$), and it is seen that metal usage can be significantly reduced without any serious impairment on the Q-factor. However, the cost in the Q-factor for small area coverage steeply increases.
A similar trade-off can be studied between different physical and structural parameter combinations. 

\subsection{Multiple Parameters Used}
\label{sec:MultiWeights}
This section examines the effect of multiple regularity parameters used simultaneously. The multi-criteria optimization problem reads
\begin{equation}
\begin{split}
\underset{\Gvec}{\T{minimize}} \quad & \left\{\frac{Q}{Q_\T{lb}}, r_\T{hom},r_\T{slot}\right\},\\
\end{split}
\label{eq:TopoOptMultiWeight}
\end{equation}
where the criteria consider normalized Q-factor $Q/Q_\T{lb}$, the homogeneity of the material~$r_\T{hom}$ and the occurrence of infinitesimal slots~$r_\T{slot}$.

\begin{figure}
\centering
\includegraphics[width=0.48\textwidth]{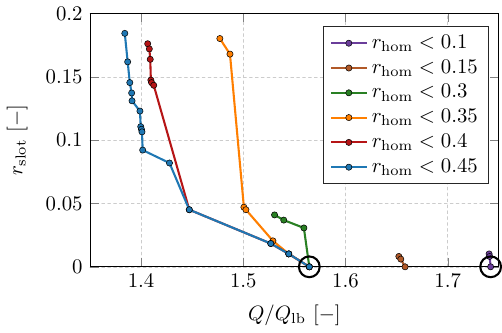}
\caption{Trade-off between normalized Q-factor and the number of slots in the structure. Homogeneity parameter $r_\T{hom}$ is used as a curve parametrization. Two black circles highlight the position of shapes shown in Fig.~\ref{pic:TopoOptMultiWeight}.}\label{pic:HomogeneitySlotTradeOff}
\end{figure}

Figure~\ref{pic:HomogeneitySlotTradeOff} shows the Pareto-frontier between all three parameters. The curves are parametrized by the homogeneity threshold. It can be seen that the infinitesimal slots can be removed from the design, albeit at the cost of a slight increase in the Q-factor. Homogeneity has a more pronounced effect on the Q-factor and should be minimized with care. It is also worth mentioning that there is a correlation between homogeneity and the presence of slots. The low number of~$r_\T{hom}$ means the slots are suppressed in the resulting structure. This is, however, not true in general as homogeneity increases, see the blue curve for~$r_\T{hom} < 0.45$ in Fig.~\ref{pic:HomogeneitySlotTradeOff}.

\begin{figure*}
\centering
\includegraphics[width=0.99\textwidth]{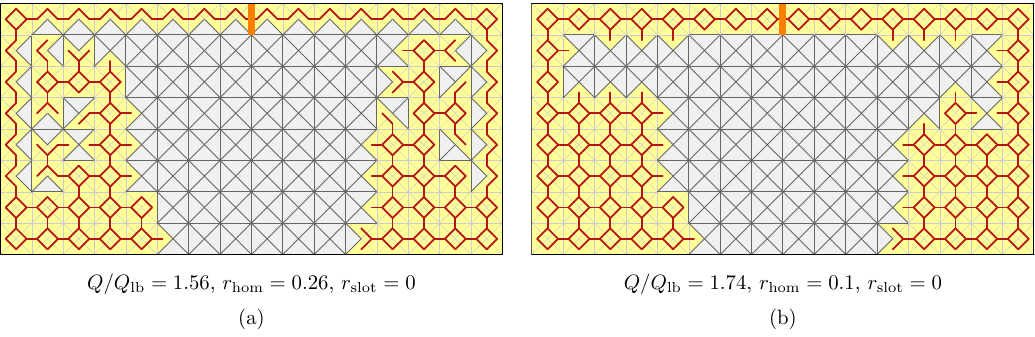}
\caption{Comparison of two structures resulting from~\eqref{eq:TopoOptMultiWeight} with different design priorities. The orange line represents the position of the discrete feeding port.}\label{pic:TopoOptMultiWeight}
\end{figure*}

The solutions obtained with two homogeneity thresholds are shown in Fig.~\ref{pic:TopoOptMultiWeight} and their respective performance is compared. Both designs have a similar overall shape and do not contain infinitesimal slots. Nevertheless, as the design in Fig.~\ref{pic:TopoOptMultiWeight}(b) varies less in material distribution (lover $r_\T{hom}$ value), it performs worse in terms of physical quantity $Q/Q_\T{lb}$ as compared to the design in Fig.~\ref{pic:TopoOptMultiWeight}(a). The final design selection is up to the designer. 

\section{Conclusion}
\label{sec:Conclusion}
This article introduces a novel regularity concept for discrete topology optimization of radiating structures. The regularity parameters mitigate known manufacturing issues, such as isolated islands, point connections, or checkerboard problems. The discretized optimization region is expressed as a graph, which leads to a faster evaluation of graph-based sparse matrices. These matrices can be used to constrain structural artifacts during the optimization process. The evaluation of regularity parameters is computationally cheap and easy to incorporate into arbitrary discrete optimization schemes. The examples illustrate the trade-offs that arise from incorporating regularity parameters into the optimization. It is observed that easier-to-manufacture designs tend to exhibit lower physical performance.

The presented approach can be adapted to introduce additional graph-based parameters that assist with inverse antenna design, incorporating prescribed structural or topological features. For example, detecting loops, finding the longest current path within the structure, or minimizing electrical size. The theory can also be expanded for three-dimensional discretization elements.

\appendices
\section{Manufacturing and Measurement}
\label{sec:ManuMeas}
\begin{figure}
\centering
\includegraphics[width=0.48\textwidth]{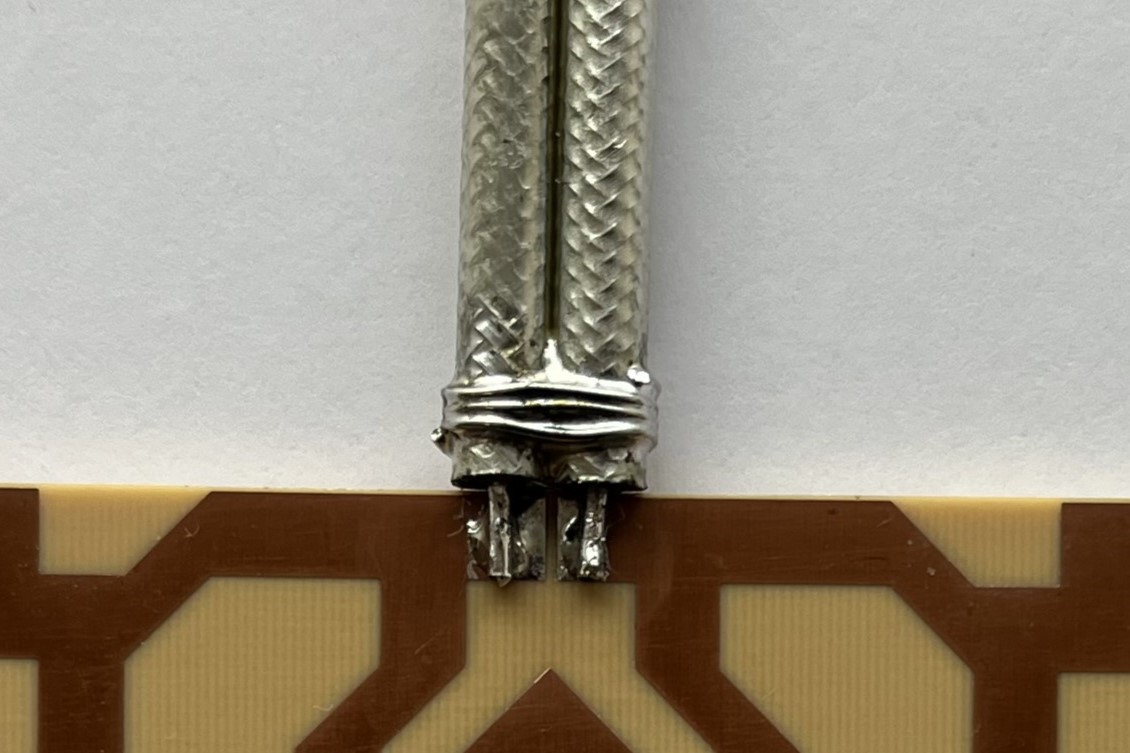}
\caption{Detail of a differential probe used as an antenna feeder.}\label{pic:AntennaFeeder}
\end{figure}

The discrete delta gap~\cite{gibson2021moments} needs to be replaced with a realistic feeder, as shown in Fig.~\ref{pic:AntennaFeeder}.
The differential probe consists of two coaxial cables with their outer conductors connected, and is used to measure the input impedance through measurement of s-parameters and subsequent transformation~\cite{739191} through an equation
\begin{equation}
\Zin = 2Z_0\frac{2 + s_{11} - s_{12} - s_{21} + s_{22}}{2 - s_{11} + s_{12} + s_{21} - s_{22}},
\end{equation}
where $Z_0 = 50\,$$\Omega$, and individual s-parameters are obtained from a two-port measurement by a Rohde \& Schwarz ZVA 40 \ac{VNA} ~\cite{dat:RHZVA40}. To further decrease the influence of antenna surroundings, the measurements were made in an anechoic chamber.

\begin{figure}
\centering
\includegraphics[width=0.48\textwidth]{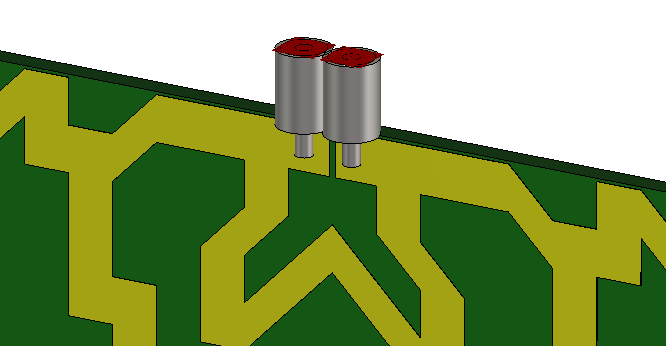}
\caption{Detail of a simulation model of a differential probe used as an antenna feeder.}\label{pic:ModeledPort}
\end{figure}

Measured data had to be corrected by performing de-embedding, which removes the effect of the differential probe on the input impedance. The \ac{VNA} was calibrated at the reference plane at the SMA connectors of the cables used. To shift the reference plane to the end of the differential probe, a simplified model of the transition from the antenna input port to the differential probe was modeled in CST Microwave Studio, see Fig.~\ref{pic:ModeledPort}. The Open-Short-Match calibration standards~\cite{CollierSkinner2007Microwave} were connected to the end of the differential probe. The gathered data were used to perform an offline measurement calibration, correcting the measured results.

\section{Graph Matrices}
\label{sec:GraphMatrices}
Considering graph $G = (\mathscr{V},\mathscr{E})$ with $\mathscr{V}$ denoting the set of nodes and $\mathscr{E}$ standing for the edges, several matrices can be defined~\cite{book:BollobasModernGraphTheory}. Incidence matrix~$\M{M}\in\mathbb{B}^{V\times E}$ reads
\begin{equation}
M_{mn}= 
  \begin{cases}
    \displaystyle\, 1 & \text{the $m$-th vertex is incident to $n$-th edge}, \\
    \displaystyle\, 0 & \text{otherwise},
  \end{cases}
\end{equation}
adjacency matrix $\M{A}\in\mathbb{B}^{V\times V}$ reads
\begin{equation}
A_{mn} = 
  \begin{cases}
    \displaystyle 1 & \text{the $m$-th vertex is adjacent to the $n$-th vertex}, \\
    \displaystyle\boldsymbol{0} & \text{otherwise},
  \end{cases}
\end{equation}
and diagonal degree matrix $\M{D}\in\mathbb{R}^{V\times V}$ is obtained through relation
\begin{equation}
\M{D} = \Mmat\Mmat^\trans - \M{A},
\end{equation}
where elements are equal to the vertices' degrees~\cite{book:BollobasModernGraphTheory}.

The matrix used for the investigation of neighborhood homogeneity~\eqref{eq:RegMetric} is homogeneity matrix $\Hmat$, defined as
\begin{equation}
\widetilde{\Hmat} = \B{\Mmat^\trans\Mmat}
\end{equation}
where $\B{-}$ is the Boolean rounding operator~\eqref{eq:BooleanRounding}, and $\widetilde{\Hmat}$ is the auxiliary matrix used for defining the homogeneity matrix
\begin{equation}
\Hmat = \begin{bmatrix}
\dfrac{\widetilde{\Hmat}_1}{\left\Vert\widetilde{\Hmat}_1\right\Vert_1} & \hdots & \dfrac{\widetilde{\Hmat}_N}{ \left\Vert\widetilde{\Hmat}_N\right\Vert_1}
\end{bmatrix}^\trans.
\end{equation}
Focusing on the~$m$-th row of matrix~$\Hmat$, if the considered basis function has in total $k$ adjacent basis functions, then the $m$-th row has exactly $k+1$ non-zero equal elements and its norm reads $\left\Vert\Hmat_m\right\Vert_1 = 1$.

In the general case, when the optimization variable consists of discretization elements instead of the shared edges between them, the homogeneity matrix can be defined with the above matrices as
\begin{equation}
\Hmat = \left(\Dmat+\M{E}\right)^{-1}\left(\Amat+\M{E}\right),
\end{equation}
where $\M{E}$ is the identity matrix of corresponding size.

To find the point connections, we introduce two additional incidence matrices, starting with
\begin{equation}
M_{\T{nt},mn} = 
  \begin{cases}
    \displaystyle 1 & \text{the $m$-th node is incident to the $n$-th triangle}, \\
    \displaystyle 0 & \text{otherwise}
  \end{cases}
\end{equation}
assigning connections between nodes and triangles, and the incidence matrix
\begin{equation}
M_{\T{nb},mn} = 
  \begin{cases}
    \displaystyle 1 & \text{the $m$-th node is incident to the $n$-th edge}, \\
    \displaystyle 0 & \text{otherwise},
  \end{cases}
\end{equation}
which establishes the relation between discretization edges and nodes. These two matrices directly consider nodes in the discretized structure.

All matrices introduced in this appendix are sparse, implying fast numeric manipulation and operations when used.

\section{Graph Matching}
\label{sec:GraphMatching}
Given graph~$G = (\mathscr{V},\mathscr{E})$, matching~$\mathscr{M}$ in~$G$  is a set of edges that are pairwise non-adjacent and do not include any loops, meaning no two edges share a common vertex~\cite{book:BollobasModernGraphTheory}. A perfect matching is not possible in graphs with an odd number of vertices. Although  set~$\mathscr{M}$ may not be unique, its cardinality~$\vert\mathscr{M}\vert$ is unambiguous.

\begin{figure}
\centering
\includegraphics[scale=1]{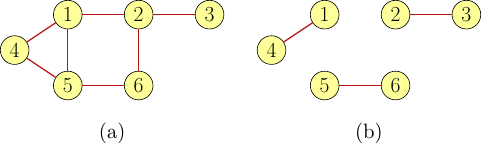}
\caption{An example of graph matching. (a) Original graph $G$. (b) Matching~$\mathscr{M}$ of graph~$G$. }\label{pic:GraphMatching}
\end{figure}

An example of graph matching is shown in Fig.~\ref{pic:GraphMatching}. It is clearly visible that multiple solutions are possible. However, there will always be only three enabled edges which correspond with the number of vertices divided by two~\cite{book:BollobasModernGraphTheory}.

\bibliographystyle{IEEEtran}
\bibliography{library}

\end{document}